\documentclass[pra, 10pt, letterpaper,oneside,notitlepage, twocolumn]{revtex4-1}
\usepackage[latin1]{inputenc}
\usepackage{mathtools}
\usepackage{amsfonts}
\usepackage{array}
\usepackage[T1]{fontenc}
\usepackage{amssymb}
\usepackage[left=0.5in, right=0.5in, top=1in, bottom=1in]{geometry}

\usepackage{graphicx}
\usepackage{float}
\newcommand{\bibs}{C:/Users/seanm_000/Dropbox/References/BibFile}

\begin{document}
\title{Dalgarno-Lewis perturbation theory for nonlinear optics}
    \date{\today}
    \author{Sean Mossman}
    \author{Rick Lytel}
    \author{Mark G. Kuzyk}
    \affiliation{Department of Physics and Astronomy, Washington State University, Pullman, Washington  99164-2814}

    \begin{abstract}
    We apply the quadrature-based perturbation method of Dalgarno and Lewis to the evaluation of the nonlinear optical response of quantum systems. This general operator method for perturbation theory allows us to derive exact expressions for the first three electronic polarizabilities which require only a good estimate of the ground state wave function, makes no explicit reference to the underlying potential, and avoids complexities arising from excited state degeneracies. We apply this method to simple examples in 1D quantum mechanics for illustration, exploring the sensitivity of this method to variational solutions as well as poor numerical sampling. Finally, to the best of our knowledge, we extend the Dalgarno-Lewis method for for the first time to time-harmonic perturbations, allowing dispersion characteristics to be determined from the unperturbed ground state wave function alone.
    \end{abstract}

    \maketitle

    \section{Introduction}\label{sec:intro}
    The collection of phenomena emergent from the nonlinear coupling of electric fields is varied and unique in the amount of control they promise over the phase, polarization and frequency of light, but the efficiency of such processes is limited by the inherent response of the material mediating the optical coupling\cite{boyd09.01}. Development of efficient molecular systems has progressed steadily but is hampered by the inability to accurately classify and quantify the quantum-scale characteristics which result in optimized nonlinear optical responses\cite{watki09.01, champ06.01, mochi06.01, lytel13.01, chen05.01}.

    Determining the macroscopic response of a material, from nonlinear crystals to organic dye-doped polymers, is a complicated process which must take many contributions into account. For applications which require the fastest dynamic response, the material efficiency is fundamentally limited by the response characteristics of the microscopic electronic systems which make up the material. The determination of fundamental characteristics which permit electronic quantum systems to exhibit maximal nonlinear responses are of primary interest.

    Such studies are often carried out by using the sum-over-states (SOS) method for describing the perturbative response of a structure to an external optical field\cite{orr71.01}. As an extension of standard time-dependent perturbation theory, this method requires the complete set of energy eigenfunctions and energy eigenvalues for the system in question. In practice, the SOS method is often implemented with two or three states alone, introducing a systematic truncation error. The method we present here allows for an exact calculation of the nonlinear susceptibilities with knowledge of the ground state wave function alone, bypassing any complications due to difficulty in obtaining excited state solutions or identifying complex degeneracies in the energy spectrum.

    Here we present a method of calculating the nonlinear response of quantum systems by specializing the Dalgarno-Lewis (DL) perturbation theory \cite{dalga55.01,schwa59.01,mavro91.01,harri77.01,maize11.01} to the computation of the dispersion of the electronic, nonlinear optical hyperpolarizabilities in the electric dipole approximation. The DL method replaces the sum over excited states from perturbation theory with an expectation value of an operator which can be deterministically determined from the ground state wavefunction. Thus, the polarizabilities may be computed with knowledge of only the ground state wave function, independent of the underlying potential energy function. In effect, DL reduces the identification of potentially exciting new electronic system to a close examination of their ground state wave functions alone.

    Section \ref{sec:SOS} reviews the conventional SOS approach to obtaining the hyperpolarizability tensors, which constitutes a starting point for the DL method. Section \ref{sec:DL} derives the DL approach within the context of nonlinear optics, shows how it replaces the sum over states with an operator, and describes how the action of that operator on the ground state can be determined exactly. The application is extended to include the second hyperpolarizability by defining yet another DL operator.  Section \ref{sec:app} applies the DL perturbation theory to the half harmonic oscillator and the infinite slant well, using each as a platform for describing the numeric difficulties and strengths of the method, followed by a discussion of applicability of variational solutions to DL calculations. Section \ref{sec:DLres} extends the DL method to dispersion phenomenon by quantizing the photon field and incorporating the photon frequency into the perturbation theory. Section \ref{sec:conclusion} concludes with a discussion of the application of DL to practical systems and the interpretation of the DL operators in the context of virtual transitions.
    \\

    \section{Sum-over-states evaluation of hyperpolarizabilities}\label{sec:SOS}

    The SOS expressions for the polarizabilities are a result of applying standard Rayleigh-Schr\"odinger perturbation theory to the dipole perturbation\cite{orr71.01, boyd09.01, griff05.01}. Beginning with an arbitrary Hamiltonian and the corresponding set of unperturbed, electronic stationary states,
    \begin{equation}
        H_0|n\rangle = E_n|n\rangle,
    \end{equation}
    we take the dominant radiation process to be represented by an interaction potential of the form $H' = - \vec{\mu}\cdot\vec{\mathcal{E}}(t)$,
    the scalar product of the dipole moment with the external, time-harmonic field, which is taken to be uniform across the system at any given time. Thus, our perturbing potential is
    \begin{equation}
        H' = \text{Re}\left[-e\vec{r}\cdot\vec{\mathcal{E}}e^{-i\omega t}\right],
        \label{eq:perturb}
    \end{equation}
    where the electric charge $e$ and the electric field strength $\vec{\mathcal{E}}$ are treated as constant parameters.

    The response of the system to an electric field at low temperatures is determined by the dipole moment of the ground state perturbed by the electric field. The perturbed state is defined by
    \begin{equation}
        H|\Psi_0\rangle = \big( H_0 + H'\big)|\Psi_0\rangle = E'_0\Psi_0\rangle.
    \end{equation}
    The expectation value of the dipole moment in the perturbed ground state can be expressed by the expansion
    \begin{align}
    \langle \Psi_0|\mu^i|\Psi_0\rangle = &\mu_{00}^i + \alpha_{ij}\mathcal{E}^j\nonumber\\
    &+ \beta_{ijk}\mathcal{E}^j\mathcal{E}^k + \gamma_{ijkl}\mathcal{E}^j\mathcal{E}^k\mathcal{E}^l+\dots,
    \end{align}
    which separates the contributions to the total dipole moment which arise from different orders of applied electric field strength. Therefore, the resulting radiation is fully determined by the applied fields and the polarizability tensors, $\alpha_{ij}$, $\beta_{ijk}$, $\gamma_{ijkl}$, etc.

    The usual procedure for determining the linear polarizability $\alpha_{ij}$ and the first and second hyperpolarizabilities, $\beta_{ijk}$ and $\gamma_{ijkl}$ respectively, begins with time-dependent perturbation theory. The perturbation given by Eq. \ref{eq:perturb} can be taken in two parts: the time-harmonic electric field with a constant magnitude and the dipole operator. The time-harmonic part contributes frequency to the denominator then drops out of the perturbation theory leaving expressions much like time-independent perturbation theory on the position operator. This procedure results in the analytical expressions for the polarizability,
    \begin{equation}
        \alpha_{ij}(\omega) = e^2\mathcal{P_F}\sum_{m\neq0}\left(\frac{x^i_{0m}x^j_{m0}}{E_{m0}-\hbar\omega}\right),
        \label{eq:sosalpha}
    \end{equation}
    and the hyperpolarizabilities,
    \begin{equation}
        \beta_{ijk}(-\omega_\sigma;\omega_1,\omega_2) = \frac{e^3}{2}\mathcal{P_F}\sum_{m,n \neq 0}\left(\frac{x^i_{0n}(x^j_{nm}-x^j_{00}\delta_{nm})x^k_{m0}}{(E_{m0}-\hbar\omega_\sigma)(E_{n0}-\hbar\omega_2)}\right),
        \label{eq:sosbeta}
    \end{equation}
    and
    \small
    \begin{align}
        &\gamma_{ijkl}(-\omega_\sigma;\omega_1,\omega_2,\omega_3)\hspace{4cm}\nonumber\\
        &= \frac{e^4}{6}\mathcal{P_F}\left(\sum_{m,n,p \neq 0} \frac{x^i_{0n}(x^j_{nm}-x^j_{00}\delta_{nm})(x^k_{mp}-x^k_{00}\delta_{mp})x^l_{p0}} {(E_{n0}-\hbar\omega_\sigma)(E_{m0}-\hbar\omega_2-\hbar\omega_3)(E_{p0}-\hbar\omega_3)}\right.\nonumber\\
        &\hspace{0.5cm}-\left.\sum_{n,m \neq 0}\frac{x^i_{0n}x^j_{n0}x^k_{0m}x^l_{m0}} {(E_{n0}-\hbar\omega_\sigma)(E_{n0}-\hbar\omega_2-\hbar\omega_3)(E_{m0}-\hbar\omega_3)}\right),
        \label{eq:sosgamma}
    \end{align}
    \normalsize
    where $\mathcal{P_F}$ indicates a sum over all tensor  component permutations, $x^i_{nm} = \langle n|x^i|m \rangle$ for the $i$th Cartesian component of the position operator, $E_{n0} = E_n-E_0$, and the sums include all unperturbed \emph{excited} states of the system. The problem of secular divergences, centering around the inclusion of the ground state in the above sums, has been a long discussed issue and has been settled from a few different perspectives \cite{orr71.01, bisho94.01}.

    \begin{figure}
        \includegraphics{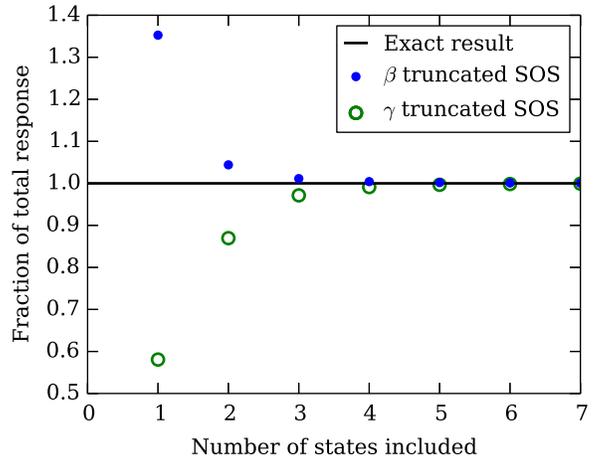}
        \caption{The convergence behavior of the SOS calculation for the half harmonic oscillator. Typical SOS calculations have similar behavior.}
        \label{fig:CHOSOSstates}
    \end{figure}

    SOS expressions are fundamental to all of perturbation theory and show up in a plethora of applications, from determining Van der Waals coefficients for interatomic and intermolecular forces to multiphoton ionization cross-sections. It is commonplace to see that, when this perturbation theory is correctly applied, the sums converge sufficiently after including a small number of excited states. However, in practice, one must always truncate the sum at some point. A two or three level model might be all that is possible with computational results, and highly excited states are often particularly difficult to include. Furthermore, the number of terms in th sum for $\beta$ scales as the square of the number of states included while $\gamma$ scales as the cube; the DL calculations scale linearly, each order requiring only one additional integration, and can therefore be completed significantly faster, in principle.

    \section{Dalgarno-Lewis perturbation formalism in the static field limit}\label{sec:DL}

    The DL perturbation method recasts the products of transition moments and energy denominators from Eqs. \ref{eq:sosalpha}-\ref{eq:sosgamma} as operators in such a way that the sum-over-states can be removed\cite{mavro91.01}.  The SOS expressions are replaced by ground state expectation values of operator products containing all of the physics of the perturbations.  In this section, we show how this is accomplished and derive explicit expectation values for the nonlinear optical susceptibilities in the static field limit, leaving dispersion for Section \ref{sec:DLres}.

    \subsection{First hyperpolarizability}\label{sec:betaDL}

    Suppose that an operator $F^i$ exists such that the following commutator equation holds
    \begin{equation}
        [F^i, H_0] = (x^i - x^i_{00}),
        \label{eq:Fdef}
    \end{equation}
    where $H_0$ is the unperturbed Hamiltonian and the right hand side represents the perturbation potential discussed in Sec. \ref{sec:SOS} up to constant parameters. Eq. \ref{eq:Fdef} then defines the off-diagonal matrix elements of the operator $F^i$ as
    \begin{equation}
        \langle m|F^i|n\rangle = \frac{\langle m |x^i| n\rangle}{E_n-E_m}\text{\ \ for $m\neq n$.}
        \label{eq:Fmatrixelem}
    \end{equation}
    Substituting Eq. \ref{eq:Fmatrixelem} into Eqs. \ref{eq:sosbeta} yields
    \begin{equation}
        \beta_{ijk} = \frac{e^3}{2}\mathcal{P_F}\sum_{m,n\neq0}\langle 0|-F^i|n \rangle\langle n|x^j-x^j_{00}|m \rangle\langle m|F^k|0 \rangle.
    \end{equation}

    The utility of this operator approach then becomes clear: We have decoupled the $n$ and $m$ indices allowing the sums to be removed by way of closure. Without loss of generality we may take $\langle 0|F|0 \rangle=0$ and therefore Eqs. \ref{eq:sosalpha} and \ref{eq:sosbeta} become
    \begin{equation}
        \alpha_{ij} = e^2\mathcal{P_F}\langle 0|F^ix^j|0 \rangle
        \label{eq:DLalphaOp}
    \end{equation}
    and
    \begin{equation}
        \beta_{ijk} = -\frac{e^3}{2}\mathcal{P_F}\langle 0|F^i(x^j-x^j_{00})F^k|0 \rangle.
        \label{eq:DLbetaOp}
    \end{equation}
    The sum-over-states has been reduced to the computation of an expectation value of a single product of operators, with the operators $F^i$ to be determined for each Cartesian direction.

\subsection{Properties of the DL operator $F^i$} \label{sec:propF}
    From inspection of the defining equation for $F^i$, Eq. \ref{eq:Fdef}, it is clear that $F^i$ must be an anti-Hermitian operator, uniquely determined except for the addition of any operator that commutes with $H_0$, which would leave Eq. \ref{eq:Fdef} invariant. We would like to project this operator into position space but any local spatial representation of such an operator would necessarily be imaginary, any function position must be imaginary to also represent an anti-Hermitian operator. This is in conflict with the SOS representation, as can be seen more explicitly by directly representing the operator $F^i$ in terms of its matrix elements as
    \begin{equation}
        F^i = \sum_{n\neq m} \left(\frac{\langle n|x|m \rangle}{E_m-E_n} |n\rangle\langle m|\right),
        \label{eq:Fsos}
    \end{equation}
    where for real transition moments, the matrix elements of $F^i$ are similarly real. To determine all of the matrix elements of the operator $F^i$ requires the completion of the sum-over-states as dictated by Eq. \ref{eq:Fsos}.

    However, to compute the nonlinear response of a system at low temperature we are interested only in the nonlinear coupling of the ground state through virtual transitions back to the ground state. Therefore, the only information we require is the action of the $F^i$ operator on the ground state, and this specific operation is expressible as a function of $\vec{x}$. To see this, we define a function $F_0^i(\vec{x})$ such that
    \begin{equation}
        \langle \vec{x}|F^i|0 \rangle \equiv F_0^i(\vec{x})\psi_0(\vec{x})
        \label{eq:Ffunctiondef}
    \end{equation}
    and, reflecting the anti-Hermitian nature of the operator, we also require
    \begin{equation}
        \langle 0|F^i|\vec{x} \rangle = -F_0^{i*}(\vec{x})\psi^*_0(\vec{x}).
        \label{eq:Fadjfunctiondef}
    \end{equation}
    This procedure is always valid as the state vector which results from the operation of $F^i$ on the ground state is also a state vector in the same Hilbert space, as is evident in Eq. \ref{eq:Fsos}, and its projection into spacial coordinates can be represented by some function of $\vec{x}$, which we choose to write as $F_0^i(\vec{x})\psi_0(\vec{x})$. It is important to realize that $F_0^i(\vec{x})$ is not the position representation of the operator $F^i$, but rather a function that represents the operation of $F^i$ on the ground state specifically.

    In order to calculate the function $F^i_0(\vec{x})$, we return to the commutator definition of $F^i$, Eq. \ref{eq:Fdef}. Taking a mechanical Hamiltonian of the form $H_0 = \frac{p^2}{2m} + V_0$ and projecting into spatial coordinates,
    \begin{equation}
        \langle \vec{x} |[F^i,H_0]|0\rangle = \langle \vec{x}|x-x_{00}|0\rangle,
        \label{eq:commmatrixelem}
    \end{equation}
    yields
    \begin{align}
        \frac{\hbar^2}{2m}\Big(\psi_0(\vec{x})\nabla^2F^i_0(\vec{x})+2\nabla &F^i_0(\vec{x})\cdot\nabla\psi_0(\vec{x})\Big)\nonumber\\ &=(x^i-x^i_{00})\psi_0(\vec{x}),
        \label{eq:Fdiff}
    \end{align}
    or equivalently
    \begin{align}
    \frac{\hbar^2}{2m}\frac{1}{\psi_0(\vec{x})} \nabla\cdot\left(\psi_0^2(\vec{x})\nabla F_0(\vec{x})\right) = \left(x-x_{00}\right)\psi_0(\vec{x}),
    \label{eq:Fdiff2}
    \end{align}
    which can be solved in general to obtain the function $F_0^i(\vec{x})$ with knowledge of the ground state wave function alone. Eqs. \ref{eq:DLalphaOp} and \ref{eq:DLbetaOp} can then be calculated by integrating across products of $F_0^i(\vec{x})$, $\psi_0(\vec{x})$, and $\vec{x}$.

    For a Cartesian separable system, that is $\psi_0(\vec{x}) = \prod_i\phi_0^i(x^i)$, the integral solution to Eq. \ref{eq:Fdiff} is
    \begin{align}
        F^i_0(x^i) = \frac{2m}{\hbar^2}\int^{x^i} &\frac{1}{\left(\phi^i_0(x')\right)^2}\label{eq:Fintsol}\\
        &\left[ \int_a^{x'}(\xi^i - \xi^i_{00}) \left(\phi^i_0(\xi)\right)^2 d\xi\right]dx' + C,\nonumber
    \end{align}
    where $a$ is chosen such that $\lim_{x\rightarrow a}\phi^i_0(x)=0$ to satisfy the boundary conditions on the differential equation and $C$ is undetermined but chosen such that $\langle 0|F|0 \rangle = 0$ for convenience. The ground state has the important characteristic of being nodeless except for two points, at most, which necessarily bound the domain of the state function. Interestingly, the function $F^i_0(\vec{x})$ must contain phase information from $\psi_0(\vec{x})$, hence we see the square of the wave function, rather than the absolute square.

    Summarizing, the DL formalism starts with the definition of a set of anti-Hermitian operators $F^i$ via a commutation relation with the unperturbed Hamiltonian and results in an explicit expression for the spatial representation of the operator's inner product with the ground state in terms of quadratures, which depend only on the ground state wave function. If the ground state is known, then the functions $F^i_0(\vec{x})$ are calculable.  With the functions $F^i_0(x)$ in hand, one for each Cartesian direction,  Eqs. \ref{eq:DLalphaOp} and \ref{eq:DLbetaOp} may be projected into position space and integrated to obtain the linear and first hyperpolarizability tensors, regardless of the symmetries of the system or the degeneracies in the spectrum. The ground state alone determines the response of the system.

\subsection{Second hyperpolarizability}\label{sec:gammaDL}
    The second hyperpolarizability $\gamma_{ijkh}$ requires an additional integration to calculate the response while still only requiring the ground state wave function. Substituting Eq. \ref{eq:Fmatrixelem} into Eq. \ref{eq:sosgamma} we obtain
    \begin{align}
        \gamma_{ijkh} = \frac{e^4}{6}\mathcal{P_F}\Bigg[\sum_{n \neq 0}&\left(\frac{\langle 0|F^i(x^j-x^j_{00})|n \rangle\langle n|(x^k-x^k_{00})F^h|0\rangle} {E_{n}-E_0}\right)\nonumber\\
        &-\langle 0|F^iF^j|0\rangle\langle 0|x^kF^h|0\rangle\Bigg],\label{eq:gammainit}
    \end{align}
    which must be further simplified by defining a new operator $G^{ij}$ in analogy with $F^i$:
    \begin{equation}
        [G^{ij},H_0] = (x^i-x^i_{00})F^j - \langle x^iF^j\rangle_{00}.
        \label{eq:Gdef}
    \end{equation}
    As with $F_0^i(\vec{x})$, the matrix elements are given by
    \begin{equation}
        \langle n|G^{ij}|0 \rangle = -\frac{\langle n|(x^i-x^i_{00})F^j|0 \rangle}{E_n-E_0}.
    \end{equation}
    The last sum can then be removed from Eq. \ref{eq:gammainit} using completeness, resulting in the operator expression
    \begin{align}
        \gamma_{ijkl} = \frac{e^4}{6}\mathcal{P_F}&\Big( \langle 0|F^i(x^j-x^j_{00})G^{kl}|0\rangle\nonumber \\
        &- \langle 0|F^iF^j|0\rangle\langle 0|x^kF^l|0\rangle\Big).\label{eq:DLgammaOp}
    \end{align}

    Defining the spatial representation of the $G^{ij}$ operator's action on the ground state through
    \begin{equation}
        \langle \vec{x}|G^{ij}|0\rangle \equiv G_0^{ij}(\vec{x})\psi_0(\vec{x})
    \end{equation}
    and projecting Eq. \ref{eq:Gdef} into position space yields a differential equation for $G_0^{ij}(\vec{x})$:
    \begin{align}
        \frac{\hbar^2}{2m}\Big(\psi_0(\vec{x})&\nabla^2G^{ij}_0(\vec{x})+2\nabla G^{ij}_0(\vec{x})\nabla\psi_0(\vec{x})\Big)\nonumber\\
         &= \left[(x^i-x^i_{00})F^j_0(\vec{x})- \langle x^iF^j\rangle_{00}\right] \psi_0(\vec{x}).
        \label{eq:Gdiff}
    \end{align}
    Just as in Eq. \ref{eq:Fdiff}, Eq. \ref{eq:Gdiff} may be numerically solved for any two indices $i$ and $j$.  Explicit integrals may be expressed for the diagonal operators for spatially separable solutions:
    \begin{align}
        G^{ii}_0(x^i) &= \frac{2m}{\hbar^2}\int^{x^i} \frac{1}{\left(\phi^i_0(x')\right)^2}
        \label{eq:Gintsol}\\
        &\left[ \int_a^{x'}\left[(\xi^i - \xi^i_{00})F^i_0(\xi) - \langle x^iF^i\rangle_{00} \right]\left(\phi^i_0(\xi)\right)^2 d\xi\right]dx' + C,\nonumber
    \end{align}
    where $\phi^i_0(a)=0$ and the integration constant is similarly chosen such that $\langle 0|G^{ij}|0 \rangle = 0$.

    The higher order susceptibilities can be obtained by following this procedure, eliminating energy denominators by absorbing them into progressively iterative operators defined by commutation with the unperturbed Hamiltonian in analogy with Eqs. \ref{eq:Fdef} and \ref{eq:Gdef}.

\subsection{Interpretation and extension}\label{sec:interpret}
    A few interesting interpretations follow from the above formalism. The common approach to understanding the structural characteristics that underly a strong nonlinear optical response focuses on the SOS expressions. Eqs. \ref{eq:sosalpha}-\ref{eq:sosgamma} clearly indicate that the polarizabilities will be maximized for systems with strong transitions between the ground state and the lowest energy excited states, implying that one must consider the low lying collection of states to optimize the response. This problem then reduces to the problem of what underlying electronic potential could possibly generate such a collection of states. This is a difficult problem with many coupled parameters.

    However, Eqs. \ref{eq:Fdiff} and \ref{eq:Gdiff} of the DL formalism imply that the calculations of the polarizabilities are independent of the explicit form of the underlying potential and of the excited state solutions; that is to say that the entire polarization expansion can be obtained from the unperturbed ground state wave function. All information about the zero temperature perturbation theory acting on a system is contained within the ground state solution for that system.

    \begin{figure}
        \includegraphics{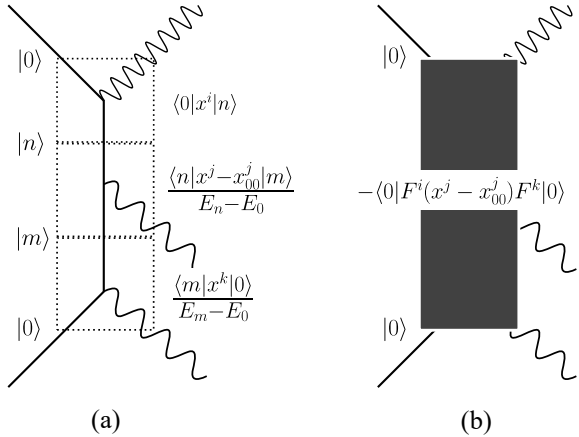}
        \caption{(a) The Feynman diagram represents a second order process where two photons interact with a quantum system in succession. The first photon causes a virtual transition from the ground state to some excited state, $|m\rangle$, the second causes a virtual transition from the state $|m\rangle$ to another excited state, $|n\rangle$, then finally the system emits a photon as it makes a final virtual transition from $|m\rangle$ back to the ground state. (b) In the DL formalism these virtual transitions are subsumed into a single expectation value. Graphically, we show a literal "black box" as a Feynman diagram in this context is rather unnecessary.}
        \label{fig:Feyn}
    \end{figure}

    An insightful point of view of the SOS expressions for nonlinear optics lies in the diagrammatic representation of virtual transitions, Feynman diagrams as shown in Fig. \ref{fig:Feyn}. Each matrix element and corresponding energy denominator in the SOS expressions can be associated with the absorption of a photon and a virtual transition of the quantum system into a sum over all excited states. From that perspective, one may take the matrix elements of the $F^i$ operator as a black box containing all of the virtual transitions, coupling the ground state with the photon field, then directly returning the system to the ground state.

    The $F^i$ operator contains all of the information for the first order perturbation theory\cite{balan10.01}. From Eq. \ref{eq:Fsos}, it is obvious that the first-order correction to the $p$th eigenstate of the unperturbed Hamiltonian can be expressed as
    \begin{equation}
        |\Psi_p^{(1)}\rangle = F^i|p\rangle = \sum_{n\neq p}\frac{\langle n|x^i|p \rangle}{E_p-E_n}|n\rangle,
    \end{equation}
    in agreement with standard perturbation theory. The $F^i$ operator explicitly determines the perturbed states, and while determining the operator $F^i$ fully would require as much work as the full sum-over-states, the first order correction to the ground state can be determined by explicitly solving for the $F^i$ operator's action on the ground state alone.

    We have explicitly projected the operator relation Eq. \ref{eq:Fdef} into spacial coordinates to yield an inhomogeneous differential equation on the function we call $F^i_0(\vec{x})$. Alternatively, one could project onto the set of atomic orbital states to recover the method of diagrammatic valence bond theory (DVB)\cite{soos89.02} for which algebraic methods have been developed\cite{ferre97.01, wlotz12.01} for tackling the numerical problems associated with many electron problems.

\section{Applications to quantum systems}\label{sec:app}

    We illustrate the application of the DL approach with two models that display the simplicity of the method and, at the same time, show how the numerical integrations must be carefully handled in order to avoid divergences.  These are expected from an examination of the Eq. \ref{eq:Fintsol} for $F^i_0$ and Eq. \ref{eq:Gintsol} for $G^{ij}_0$, where the outer integrals depend on the inverse square of the ground state wave function, which vanishes at one or both of the limits of integration.

    We apply the new method to a one dimensional half harmonic oscillator and a one dimensional slant well. All calculations of polarizabilities are normalized to the fundamental limit as a convenient choice of intrinsic units, which remove any dependence on the overall length scale\cite{kuzyk00.01, kuzyk13.01}.

    \begin{figure}
    \includegraphics{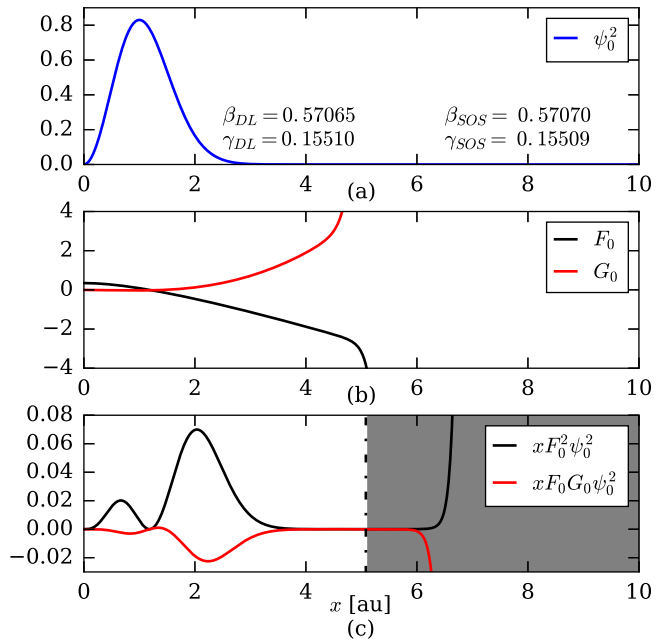}
    \caption{(a) The ground state charge density for the half harmonic oscillator and the hyperpolarizabilities calculated using DL and SOS,  (b) DL functions $F_0(x)$ and $G_0(x)$, and (c) the relevant integrands for calculation with the numerically unstable region highlighted by the grey background.}
    \label{fig:CHODL}
    \end{figure}

\subsection{Half harmonic oscillator}\label{sec:CHO}
    The Schr\"odinger equation for a half harmonic oscillator is
    \begin{equation}
        \left(\frac{\hbar^2}{2m}\frac{d^2}{dx^2}+V(x)\right)\psi_n(x)=E_n\psi_n(x)
        \label{eq:SE}
    \end{equation}
    with
    \begin{equation}
    V(x) = \begin{cases}
        \frac{1}{2}m\omega^2x^2 & x>0\\
        \infty & x\leq0
        \label{eq:CHO}
        \end{cases},
    \end{equation}
    which admits the odd solutions from the full harmonic oscillator in the positive $x$ half space. This forced asymmetry allows for a nonzero first hyperpolarizability, which can be calculated to arbitrary accuracy by way of the SOS expressions as the transition moments are analytically determined \cite{perez04.01}. For convenience we choose $m,\ e,\ \hbar$ and $\omega$ as unity.

    To determine the applicability of the DL method to numerical solutions, we approach this problem using a standard finite difference method\cite{watki10.01} for the ground state solution of the Schr\"odinger equation and then proceed with the DL integrations. Fig. \ref{fig:CHODL}a shows the charge density obtained using finite differences as well as the resulting DL hyperpolarizability results which compare very well with the exact results using analytic $x$ matrix elements and energies. Fig. \ref{fig:CHODL}b shows that the solutions for $F_0(x)$ and $G_0(x)$ become unstable but only in the region where the wavefunction has decayed to a sufficiently small magnitude such that the integrals have already converged well. Judicious choice of the integration domain allows for accurate calculations of the hyperpolarizabilities.

    One might be concerned with how lack of resolution on the ground state solution affects the DL calculation, specifically if the regions where the wave function is nearly zero affect the calculation. To investigate this, we use a generalized finite difference method which allows us to reduce the resolution within some part of the problem space to determine the affect on the results. This course resolution region is moved through the domain and the hyperpolarizabilities are calculated for each position. To mitigate anomalous numerical error resulting strictly from an abrupt change in grid resolution, we implement a smooth change of the resolution with a half cosine distribution reducing and restoring the resolution smoothly across the region.

    \begin{figure}
    \includegraphics{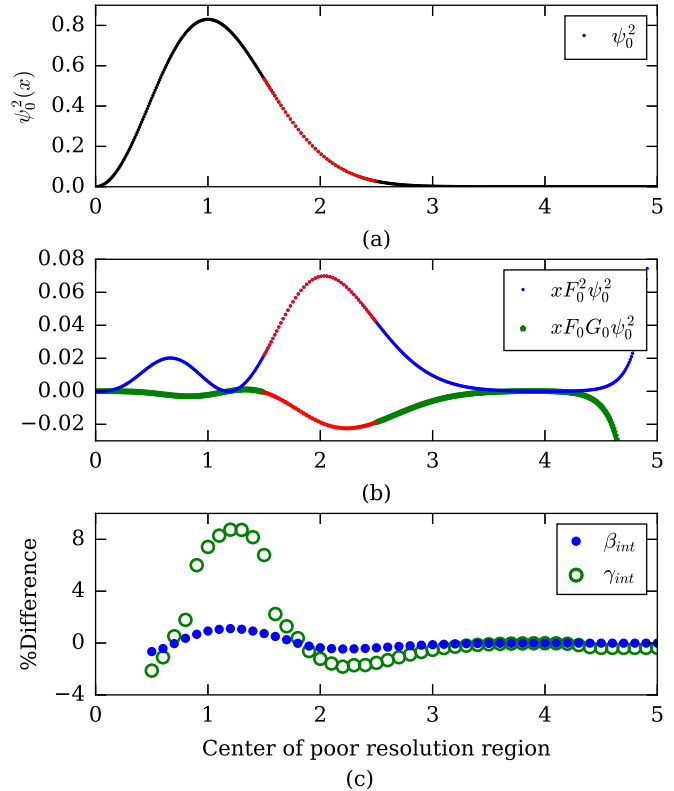}
    \caption{Resolution is reduced by a factor of two in a unit wide region centered on different points of the problem space as shown by sparser points in red, shown as an example for the center position $x=2$ for (a) the charge density and (b) the DL integrands. (c) The percent difference from the exact result as a function of the central position of the reduced resolution region.}
    \label{fig:CHODLres}
    \end{figure}

    Fig. \ref{fig:CHODLres} shows that rather than requiring high resolution in the regions where the wave function is vanishingly small, high resolution is needed where the integrand has the most curvature, just as one might expect from basic numerical integration. This result indicates that future work on systems whose solutions require a large amount of memory or a long run time can dedicate resources to those regions which require accurate integration and need not be overly concerned with the far field regions where the ground state wave function necessarily decays away.

\subsection{DL on variational ground state solutions}\label{sec:variation}

    In the previous section, we addressed how the DL solutions are affected by the resolution of the ground state wave function. Here we investigate how effectively the DL approach can approximate the polarizabilities from ground state wave functions obtained through variational methods. We choose variational functions which can only approximate the analytic results for the following 1D potentials: the half harmonic oscillator and the infinite slant well. The purpose of this discussion is not to obtain good variational solutions, but rather to observe how deviations in the ground state wave function manifest as differences in the calculated hyperpolarizability.

    The variational ground states are obtained by choosing a functional form for the ground state wave function which depends on a few parameters then varying those parameters until the expectation value of the energy is minimized for each potential. For the half harmonic oscillator potential given in Eq. \ref{eq:CHO}, we use the function
    \begin{equation}
        \psi_{var} = A\sin(\frac{\pi x}{5})\exp^{-a(x-b)^2},
        \label{eq:CHOvar}
    \end{equation}
    as opposed to the analytic solution
    \begin{equation}
        \psi_{FD} \propto x\exp\left(-\frac{1}{2}x^2\right),
    \end{equation}
    where the parameters $a$ and $b$ are varied to minimize the energy while $A$ is a normalization constant which depends on $a$ and $b$. Similarly, for the infinite slant well potential
    \begin{equation}
    V(x) = \begin{cases}
        10x & 0<x<1\\
        \infty & \text{elsewhere}
        \end{cases}
        \label{eq:ISW}
    \end{equation}
    we use the function
    \begin{equation}
        \psi_{var} = Ax^a(1-x)^b.
    \end{equation}

\begin{figure}
        \includegraphics{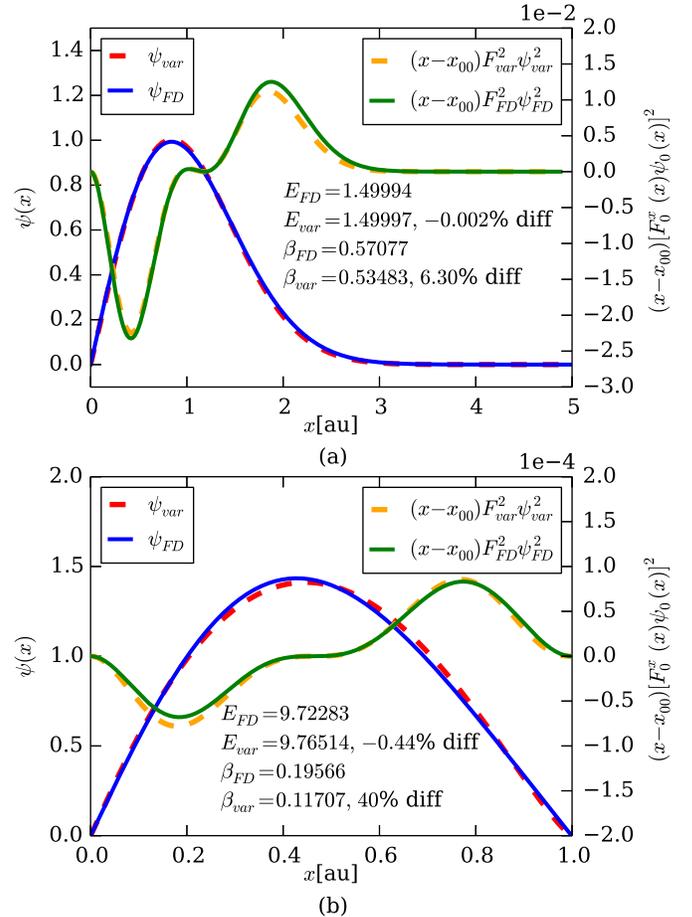}
        \caption{The variational ground state solutions and the pertinent integrand for the DL calculation of $\beta$ are plotted for (a) the half harmonic oscillator described by Eq. \ref{eq:CHO}, with $m,\ e,\ \hbar$ and $\omega=1$, and (b) the infinite slant well described by Eq. \ref{eq:ISW}.}
        \label{fig:variational}
    \end{figure}

    Fig. \ref{fig:variational} shows the ground state wave functions obtained directly by finite differences and those obtained by applying the variational principle as well as the resulting DL calculations for comparison. Recall from Eq. \ref{eq:DLbetaOp} that the relevant integrand for calculating the hyperpolarizability $\beta$ is $x\left(F_0^x(x)\psi_0(x)\right)^2$, so we should expect deviations in the wave functions to propagate through to the hyperpolarizability result accordingly.

    We observe that a variational solution can be used to effectively calculate the hyperpolarizability, but small deviations in the ground state wave function do result in appreciable deviations in the hyperpolarizability. Comparison of direct finite difference and variational methods show that the hyperpolarizabilities calculated using the DL method from the variational solutions result in a percent difference at least two orders of magnitude greater than the percent difference in the energies. The hyperpolarizability limit is calculated using the first energy difference from finite differences for both cases. When Eq. \ref{eq:CHOvar} approximates the half harmonic oscillator solution, we observe that forcing a node in the approximate ground state causes significant deviation in the $F_0$ function but in a region where the wave function has decayed sufficiently to dampen the effect on the final result. This behavior is characteristic of the open boundary condition as opposed to an infinite potential boundary.

\section{Dalgarno-Lewis for finite frequency fields} \label{sec:DLres}
    To include frequency dependence in our formulation of Dalgarno-Lewis perturbation theory we must revisit the form of our Hamiltonian as well as include the photon field in our state vectors. For the sake of clarity, we will assume that the unperturbed Hamiltonian is one dimensional, but as before, the operators can be generalized to tensor operators for higher dimensional models. With a photon field present, the unperturbed Hamiltonian is given by
    \begin{equation}
        \widetilde{H}_0 = H_0 + \sum_\omega \hbar\omega a_\omega^\dagger a_\omega
    \label{eq:Hfield}
    \end{equation}
    and the perturbing potential is given by
    \begin{equation}
        H' = (x-x_{00})\left(a_\omega  - a_\omega^\dagger \right),
    \end{equation}
    where the operators $a_\omega$ and $a_\omega^\dagger$ are the annihilation and creation operators for the photon field of frequency $\omega$, respectively, and we have suppressed the time harmonic oscillation as well as the electric field magnitude, as we would normally. Physically, each dipole transition of the system is accompanied by a single photon emission or absorption; the nonzero transition elements for a photon absorption will have the form $\langle n,N_\omega-1|x\cdot a_\omega|0, N_\omega \rangle$.

    Now, by analogy with Eq. \ref{eq:Fdef}, we define the frequency dependent $F$ operator by the commutator
    \begin{equation}
        [F^\omega,\widetilde{H}] = (x-x_{00})(a_\omega-a_\omega^\dagger)
        \label{eq:Fdeffreq}
    \end{equation}
    and consider the matrix element
    \begin{align}
    \langle n, N_\omega-1|&[F^\omega,\tilde{H}_0]|0, N_\omega \rangle \nonumber \\
    &= \langle n, N_\omega-1|(x-x_{00})\left(a_\omega  - a_\omega^\dagger\right) |0, N_\omega \rangle\\
    \langle n, N_\omega-1|&F^\omega|0, N_\omega \rangle = -\frac{x_{0n}\sqrt{N_\omega}}{E_n - E_0 - \hbar\omega}
    \end{align}
    recalling that the electric field strength is proportional to $\sqrt{N_\omega}$. Notice that $F^\omega$ operates on the state vectors of the field much like a creation or annihilation operator.

    Now, we make the following definition for the Dalgarno-Lewis function
    \begin{align}
    \langle x, M|F^\omega|0, N\rangle &=  F_0(\omega;x)\psi_0(x)\langle M|(a_\omega - a_\omega^\dagger)|N\rangle\\
    &=
    \begin{cases}
        \sqrt{N}F_0(\omega;x)\psi_0(x) \text{ for }M=N-1\\
        0\text{ otherwise}
    \end{cases}\nonumber
    \end{align}
    in analogy with Eq. \ref{eq:Ffunctiondef}. Recall that the function $F_0(x)$ is defined by the $F$ operator's action on the ground state, thus $N$ must be decreased when operated upon by $F$. Then, going back to the definition of the operator $F$ we consider the matrix elements of the defining commutation relation
    \begin{align}
        \langle x, N_\omega-1|[F, \tilde{H}_0]|0, N_\omega\rangle &= \langle x, N_\omega-1|\bar{x}(a_\omega+a^\dagger_\omega)|0, N_\omega\rangle
    \end{align}
    and after operations similar to those between Eqs. \ref{eq:commmatrixelem} and \ref{eq:Fdiff} we have
    \begin{align}
    \frac{\hbar^2}{2m}\bigg(\nabla^2F_0(\omega;x)&\psi_0(x)+2\nabla F_0(\omega;x)\cdot\nabla \psi_0(x) \nonumber\\ &+\left.\frac{2m\omega}{\hbar}F_0(\omega;x)\psi_0(x)\right) = \bar{x}\psi_0(x)
    \label{eq:DLDiffEqFreq}
    \end{align}
    or equivalently,
    \begin{align}
        \frac{\hbar^2}{2m}\nabla\cdot\left(\psi^2_0(x)\nabla F_0(\omega;x)\right) = \big(\bar{x}-\hbar\omega F_0(\omega;x)\big)\psi^2_0(x)
    \end{align}
    as defining differential equations for the frequency dependent Dalgarno-Lewis function. Unfortunately, as opposed to the static case, this differential equation is truly of second order and can not be cast into an integral equation, even in one dimension.

\subsection{Boundary conditions on the DL differential equations}
    At this point it is necessary to go back and investigate the boundary conditions on these differential equations more closely. The boundary conditions which determine $F_0(x)$ are inherited from the boundary conditions which determine the ground state wavefunction along with the physical requirement that the perturbation theory converge.

    The boundary conditions on the Schr\"odinger equation require the bound state solutions to be square integrable. For the ground state of a single electron system we know this to mean that the solution will have at most two zeros and they will bound the space over which the ground state exists.

    Going back to Eq. \ref{eq:Fdiff}, we consider the limit as $x$ approaches $a$ for $\psi_0(a) = 0$. If $\nabla\psi_0(a)\neq0$ then we recover the simple boundary condition $\nabla F(a) = 0$, as would be the case for any system which has a hard wall boundary. Otherwise, we obtain the boundary condition \begin{equation}
        \nabla F_0(a) = \frac{m}{\hbar^2}\lim_{\vec{x}\rightarrow a}\frac{(x-x_{00})\psi_0(\vec{x})}{\nabla \psi_0(\vec{x})}
    \end{equation}
    which agrees with the Eq. \ref{eq:Fintsol} condition of taking the lower bound on the interior integral to be $a$
    , shown by taking
    \begin{align}
         \lim_{x\rightarrow a} \left[ \frac{dF_0(x)}{dx} = \frac{2m}{\hbar^2} \frac{1}{\psi_0^2(x)}\int_{a}^{x}dx'(x'-x_{00})\psi_0^2(x')\right].
    \end{align}
    One boundary condition is sufficient as Eq. \ref{eq:Fdiff} is effectively a first order differential equation on $\nabla F_0(x)$ and the resulting perturbation theory is invariant upon addition of a constant to $F_0(\vec{x})$.

    In the time harmonic case, Eq. \ref{eq:DLDiffEqFreq} is truly second order and both undetermined constants contribute nontrivial functions of $\vec{x}$. To fix both undetermined coefficients, we must take both endpoints of the ground state and enforce Neumann conditions on the $F_0^\omega(x)$ function. Following similar arguments as above, we obtain the boundary condition
    \begin{equation}
        \nabla F_0(a) = \frac{m}{\hbar^2}\lim_{x\rightarrow a} \frac{(x-x_{00})\psi_0(x)}{\nabla \psi_0(x)}-\frac{\hbar\omega F_0(x)\psi_0(x)}{\nabla \psi_0(x)}.
    \end{equation}
    Again, if $\nabla \psi_0(a)=0$ then we recover the simple boundary conditions $\nabla F_0(a)=0$ but otherwise our boundary condition is itself a differential equation on $F_0(x)$.

    For practical purposes, one can always approximate a ground state solution which decays exponentially at infinity as being confined to some large box. Infinite wall boundary conditions allow one to simplify the boundary conditions to $\nabla F(\vec{a})=0$, where $a$ is taken at both nodes of the approximate wave function.

\subsection{Dispersion of the square well polarizability}
    To illustrate the validity of the dispersive DL result above, we calculate the dispersion of the polarizability for an infinite square well in one dimension of unit length with the ground state solution
    \begin{equation}
        \psi_0(x) = \sqrt{2}\sin(\pi x).
    \end{equation}
    Our task is then to solve the differential equation defined by Eq. \ref{eq:DLDiffEqFreq} treating the frequency as a parameter. The infinite square well is particularly convenient as it has a simple functional form for the ground state wave function and has distinct nodes at the edges of the well.

    \begin{figure}
    \includegraphics{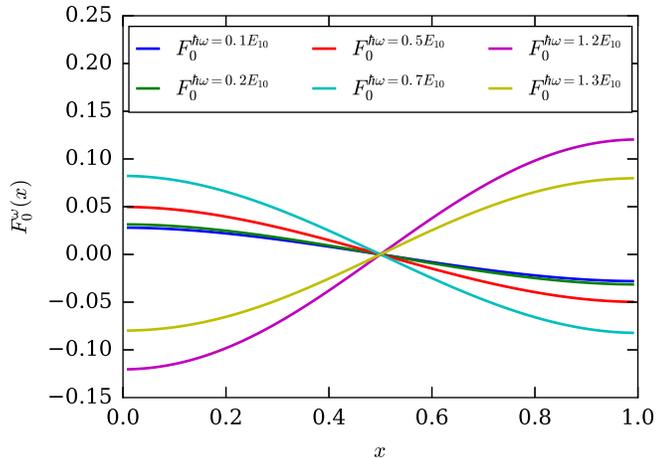}
    \caption{$F_0^\omega(x)$ DL function for the infinite square well for a variety of frequencies. As the frequency approaches a resonance, the magnitude of the $F_0^\omega$ function grows until diverging and switching signs on resonance.}
    \label{fig:Fsquarewell}
    \end{figure}

\begin{figure}
    \includegraphics{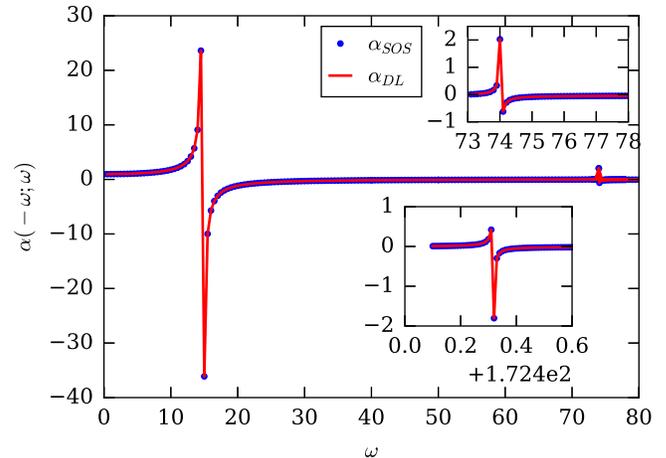}
    \caption{The dispersion curves for $\alpha$ of an infinite square well comparing the DL method to the SOS method. The first three resonances are shown.}
    \label{fig:alphaDispDL}
    \end{figure}

    Taking atomic units and the boundary conditions $F'_0(0)=0$ and $F'_0(1)=0$ we obtain the solution for the square well
    \begin{align}
        F_0^\omega(x) = & \frac{1}{2\left(1-\mathrm{e}^{\sqrt{-\pi^2-2\omega}}\right)\omega^2}\left[ \left(1-\mathrm{e}^{\sqrt{-\pi^2-2\omega}}\right)(2x-1)\omega \right. \nonumber\\ &-2\left(1-\mathrm{e}^{\sqrt{-\pi^2-2\omega}}\right)\pi\cot(\pi x) \nonumber\\
        &+ \left.2\left(\mathrm{e}^{x\sqrt{-\pi^2-2\omega}}-\mathrm{e}^{(1-x)\sqrt{-\pi^2-2\omega}}\right)\pi \csc(\pi x)\right],
    \end{align}
    shown in Fig. \ref{fig:Fsquarewell} for a variety of frequencies. For each frequency, $F_0^\omega(x)$ is integrated according to Eq. \ref{eq:DLalphaOp}, namely
    \begin{align}
        \alpha(-\omega;\omega) = \frac{e^2}{2}\int_0^L dx \Big(&-F_0^\omega(x)\psi_0^*(x) x \psi_0(x)  \nonumber \\ &-F_0^{-\omega}(x)\psi_0^*(x) x \psi_0(x)\Big),
    \end{align}
    where we have been careful to evaluate the permutation operator and introduce minus signs corresponding to the $F$ operator acting to the left on the ground state. Fig. \ref{fig:alphaDispDL} shows that the two methods agree very well, capturing at least the first three resonances, their values corresponding to $E_{n0} = \pi^2(2n^2-1/2)$, though this information was not supplied to the DL calculation in any way beyond solving Eq. \ref{eq:DLDiffEqFreq} with the ground state wave function. Evidently, the ground state contains all the information about the dispersion and amplitude of all optical transitions, retrievable through a few integrations. One could also determine the unperturbed energy eigenvalues from the position of the resonances. While this example has been integrated analytically, the power in the DL method is the ease with which numerical methods can be brought to bear on integration of a single function, as opposed to determining many eigenfunctions, integrating many matrix elements, and completing the necessary sums.

\section{Conclusions}\label{sec:conclusion}

    We have shown here that the application of the Dalgarno-Lewis perturbation formalism to the calculation of nonlinear optical coefficients can provide a high degree of accuracy for simple, single electron systems, with many advantages over the customary SOS expressions. The remarkable feature of the DL perturbation method is that it requires only the action of the DL operators \emph{on the ground state of the system} and that action is calculable from quadratures of the ground state wave function. The entire SOS expression, which requires knowledge of all states and energies of a complex system, is replaced by a new expression that depends only on the shape of the ground state. Knowledge of neither the full excitation spectrum nor the underlying potential energy is necessary for calculating the response of the system.

    Interestingly, by generalizing the DL differential equation to include the photon frequency, we are able to determine the resonances and dispersion characteristics of a system from the unperturbed ground state wave function alone. However, this method does rely on an unperturbed Hermitian Hamiltonian of the mechanical form $H = p^2/2m + V(\vec{x})$. Therefore, it is difficult to incorporate the natural linewidth or other phenomenological damping into the dispersion calculation, leaving the imaginary parts of the susceptibilities out of reach.

    In the search for the fundamental characteristics of a quantum system with maximal nonlinear response, it is interesting to note that the discussion may be framed entirely in terms of the shape of the ground state wavefunction, as opposed to discussing energy spectra and transition strengths. This work shows that those discussions are in fact identical, though the two perspectives provide significantly different insights on the problem. The connection between the two can be traced back to the realization that for low temperatures, all nonzero transitions must connect back to the ground state, so all quantum information beyond the ground state cannot contribute to the result. The fact that the perturbed ground state can be expanded as a superposition of unperturbed excited states, and is done so using standard perturbation theory, gives an erroneous sense of importance to the excited states.

    Applying Dalgarno-Lewis to multielectron systems could present significant opportunity for development of additional tools for evaluating perturbative calculations on atomic, molecular, and solid state systems. Variational methods for determining ground state solutions under Hartree-Fock or density functional theory are extensively used; Dalgarno-Lewis could provide a more accurate and numerically efficient approach for determining hyperpolarizabilities and other perturbative corrections.

\section{Acknowledgements}
    We thank Doerte Blume and Yangqian Yan for an introduction to this method. SMM and MGK thank the National Science Foundation (ECCS-1128076) for generously supporting this work.

\bibliography{\bibs}

\end{document}